\documentclass[conference]{IEEEtran}
\IEEEoverridecommandlockouts
\usepackage{cite}
\usepackage{amsmath,amssymb,amsfonts}
\usepackage{algorithmic}
\usepackage{graphicx}
\usepackage{textcomp}
\usepackage{xcolor}
\def\BibTeX{{\rm B\kern-.05em{\sc i\kern-.025em b}\kern-.08em
    T\kern-.1667em\lower.7ex\hbox{E}\kern-.125emX}}

\usepackage{xcolor}

\begin{document}

%\title{Challenges in Implementing MLOps in Certified Medical Products with Multiple Agency Boundaries}
\title{MLOps Challenges in Multi-Organization Setup: Experiences from Two Real-World Cases}

\makeatletter
\newcommand{\linebreakand}{%
  \end{@IEEEauthorhalign}
  \hfill\mbox{}\par
  \mbox{}\hfill\begin{@IEEEauthorhalign}
}
\makeatother

\author{\IEEEauthorblockN{Tuomas Granlund}
\IEEEauthorblockA{%\textit{dept. name of organization (of Aff.)} \\
\textit{Solita}\\
Tampere, Finland \\
tuomas.granlund@solita.fi}\\
\and
\IEEEauthorblockN{Aleksi Kopponen}
\IEEEauthorblockA{%\textit{dept. name of organization (of Aff.)} \\
\textit{Ministry of Finance}\\
Helsinki, Finland \\
aleksi.kopponen@vm.fi}\\
\and
\IEEEauthorblockN{Vlad Stirbu}
\IEEEauthorblockA{%\textit{dept. name of organization (of Aff.)} \\
\textit{CompliancePal}\\
Tampere, Finland \\
vlad.stirbu@compliancepal.eu}
\linebreakand 
\IEEEauthorblockN{Lalli Myllyaho}
\IEEEauthorblockA{%\textit{dept. name of organization (of Aff.)} \\
\textit{Department of Computer Science}\\
University of Helsinki\\
Helsinki, Finland \\
lalli.myllyaho@helsinki.fi} \\ 
\and
\IEEEauthorblockN{Tommi Mikkonen}
\IEEEauthorblockA{%\textit{dept. name of organization (of Aff.)} \\
\textit{Department of Computer Science}\\
University of Helsinki\\
Helsinki, Finland \\
tommi.mikkonen@helsinki.fi}
}

\maketitle

\begin{abstract}
The emerging age of connected, digital world means that there are tons of data, distributed to various organizations and their databases. Since this data can be confidential in nature, it cannot always be openly shared in seek of artificial intelligence (AI) and machine learning (ML) solutions. Instead, we need integration mechanisms, analogous to integration patterns in information systems, to create multi-organization AI/ML systems. In this paper, we present two real-world cases. First, we study integration between two organizations in detail. Second, we address scaling of AI/ML to multi-organization context. The setup we assume is that of continuous deployment, often referred to DevOps in software development. When also ML components are deployed in a similar fashion, term MLOps is used. Towards the end of the paper, we list the main observations and draw some final conclusions. Finally, we propose some directions for future work.

%"Integration of AI-development process and software development processes, including continuous and federated ML, continuous deployment, system and software evolution;"
\end{abstract}

\begin{IEEEkeywords}
Artificial intelligence, AI, machine learning, ML, multi-organisation, integration, information systems, software engineering for AI/ML.
\end{IEEEkeywords}

\section{Introduction}

The emerging age of connected, digital world means that there are tons of data concerning just about every aspect of life, economy, and industry one can imagine. Data regarding transactions, transport, purchases, health, and human behavior are recorded en masse for various purposes. This stored data has various use cases, from analytical studies to predicting how the next pandemic advances.

For precision of analysis and prediction, it is essential to have access to all data, to ensure that the data is reliable, and that the data can be processed according to the needs at hand. However, for numerous reasons, this is not always the case. Most governments have, by law, ensured that municipal and citizen related information is only accessible to those officials that truly need the data. Hence, while some data is treated open -- for instance, tax records are open to public in Finland -- it is not feasible to share everything. For instance, health records are typically something that are kept private. However, derivative data from health records can be made public, so that predictions on sickness can be reliably made. In addition, companies may view data as a part of their strategic advance, and are not willing to share their data assets.

Understanding the implications of organizational boundaries such as those described above have not been widely considered in the context of artificial intelligence (AI) and machine learning (ML). Rather, work has fostered in areas where large datasets have emerged, and they have been available for use inside one organization, like Google or Amazon for instance. However, for cases such as AuroraAI, a Finnish initiative to create well-being to citizens at personal level via digital platforms \cite{auroraai}, these boundaries are an enormous obstacle for using ML techniques to their fullest potential. Even in cases where the state would own all the data, different organizations that host it may have their own privileges and responsibilities. For cases that involve different legal entities -- say private companies and non-governmental, non-profit  organizations, for example -- and their collaboration, the boundaries can become even more apparent, as governmental data might not be shareable with them.  

To improve, we need integration mechanisms for ML/AI, analogous to integration patterns in information systems \cite{hohpe2004enterprise} but applicable at the level of AI/ML features, to create multi-organization AI/ML systems. Like with information systems, there are several challenges that need to be tackled, including integration interfaces, scaling, privacy, governance, and so on.  

In this paper, we focus on integration and scaling of systems that include ML components. The setup we assume is that of continuous deployment \cite{fitzgerald2014continuous}, where new versions of the system can be rapidly deployed -- often referred to DevOps \cite{debois2011devops,bass2015devops} in software development. When also ML components are deployed in a similar fashion, term MLOps \cite{talagala2018mlops,allamlops} is used. As a concrete example of MLOps pipelines, we use Continuous Delivery for Machine Learning (CD4ML \cite{cd4ml}), which, while numerous other proposals exist, is one of the best-known practical incarnations used for continuous deployment of ML systems.

The rest of this paper is structured as follows. In Section 2, we present the background of the paper, which mainly consists of an overview of MLOps. Sections 3 and 4 form the core of the paper by presenting two case studies. First, in Section 3, we present integration challenges regarding ML features. As a case study, we use Oravizio, a medical device for orthopedists implemented as Software-as-a-Service that is shared by two organizations. Then, in Section 4, we address scaling of ML to multi-organization context. Here, we use the AuroraAI initiative mentioned above as a case study. In Section 5, we provide an brief discussion regarding our main observations and propose some directions for future research. Towards the end of the paper, in Section 6, we draw some final conclusions.

\section{Background}

%DevOps is Development + Operations amalgamated, without considering organization boundaries

Today, at the center of software development in many organizations is a toolset that allows delivering as soon as new features are available \cite{fitzgerald2017continuous}. The goal of continuous deployment is to enable continuous flow of value adding software artifacts from the  development to the actual production use with a quality assurance. A related concept, the DevOps phenomenon \cite{debois2011devops} -- amalgamation of development operations -- can be described as a set of practices whose goal is to shorten the commit feedback cycle without compromising quality \cite{bass2015devops}. At the core of DevOps culture is collaboration between the different actors, with no consideration for organizational boundaries between the different actors delivering software or running the pipeline \cite{davis2016effective}.

In both continuous delivery and DevOps, a continuous delivery pipeline is required to support the process, from code to delivery, and, even after the deployment monitoring the behavior of the system. It is important to notice that the automated pipeline is not about software going into production without any operator supervision, but rather the pipeline provides a feedback loop to each of the stakeholders from all stages of the delivery process. Moreover, as the software progresses through the pipeline, different stages can be triggered for example by operations and test teams by a click of a button.

Following the success of DevOps, it has become desirable to include machine learning (ML) components in systems that are deployed in real time. MLOps -- amalgamation of ML and operations -- refers to advocating automation and monitoring at all steps of ML system development and deployment. As with any piece of software, support is needed for system integration, testing, releasing, deployment, monitoring and infrastructure management.

To understand the challenges related to MLOps, let us first explain the steps necessary to train and deploy ML modules \cite{SWQD21}. As the starting point, data must be available for training. There are various somewhat established ways of dividing the data to training, testing, and cross-validation sets. Next, an ML model has to be selected, together with its hyperparameters. %of the model. 
The hyperparameters define various aspects of the model. For example, in the case of neural networks, hyperparameters define how many and what kind of layers a neural network has and how many neurons there are in each layer. 
The next step is training the model with the training data. During the training phase, the system is adjusted in an iterative fashion, so that the output has a good match with the ``right answers'' in the training material. This trained model can also be validated with different data. If this validation is successful -- with any criteria we decide to use -- the model is ready for deployment, similarly to any other component in a software systems. 

In the case of MLOps, monitoring ML related features is necessary, like any other feature. However, monitoring in the context of ML must take into account inherent ML related features, such as biases and drift that may emerge over time \cite{tsymbal2004problem}. In addition, there are techniques that allow improving the model on the fly, while it is being used. Therefore, the monitoring system must take these needs into account. 
To summarize, continuous deployment of ML features is a complex procedure that involves taking into account application code, model used for prediction, and data used to develop the model. Often, the model, which can be the core of the application, is just a small part of the whole software system, so the interplay between the model and the rest of the software and context is essential~\cite{sculley2015hidden}. In addition, adequate monitoring facilities are needed to ensure that the operations are taking place correctly.

%Applications that incorporate machine learning technologies are becoming popular due to their ability to build complex prediction systems. However, the process of continuous improvement is often complex and involves changes in the following areas: the application code, the model used for prediction, and the data used to develop the model. 

%Often, these areas are handled separately by software developers, data scientists and data engineers that rely on different skill sets and tool chains. For example, data engineers are focused on making the data more accessible, data scientist perform experiments for improving the data model, and the developers are worried about integrating the various technologies and releasing them to production. The lack of harmonized processes across these domains leads to delays and frictions, such as models never reaching production or deployments that are difficult to update or debug. Due to this variability, machine learning applications are more complex than traditional applications. These characteristics makes them harder to test, explain or improve.

%\subsection{General purpose MLOps pipelines}

While numerous proposals exist from different vendors, perhaps the most well-known incarnation of MLOps is Continuous Delivery for Machine Learning (CD4ML) \cite{cd4ml}. The approach formalized by ThoughtWorks for automating in an end-to-end fashion the lifecycle of machine learning applications. In CD4ML, a cross-functional team produces machine learning applications based on code, data, and models in small and safe increments that can be reproduced and reliably released at any time, in short adaptation cycles. The approach contains three distinct steps: identify and prepare the data for training, experimenting with different models to find the best performing candidate, and deploying and using the selected model in production. The work is split to an ML pipeline that works with the data, and to a deployment pipeline that deploys the result to operations (Figure \ref{fig:cd4ml}).

The above implies that there are three artifacts, in addition to those that are required by DevOps, that need version control in MLOps: (i) different data sets used for training model and their versioning; (ii) model and its versioning; and (iii) monitoring the output of the model to detect bias and other problems. 

In the following, we use these three steps and related artifacts in two MLOps related organizational challenges.

\begin{figure*}[htb]
    \centering
    \includegraphics[width=0.99\textwidth]{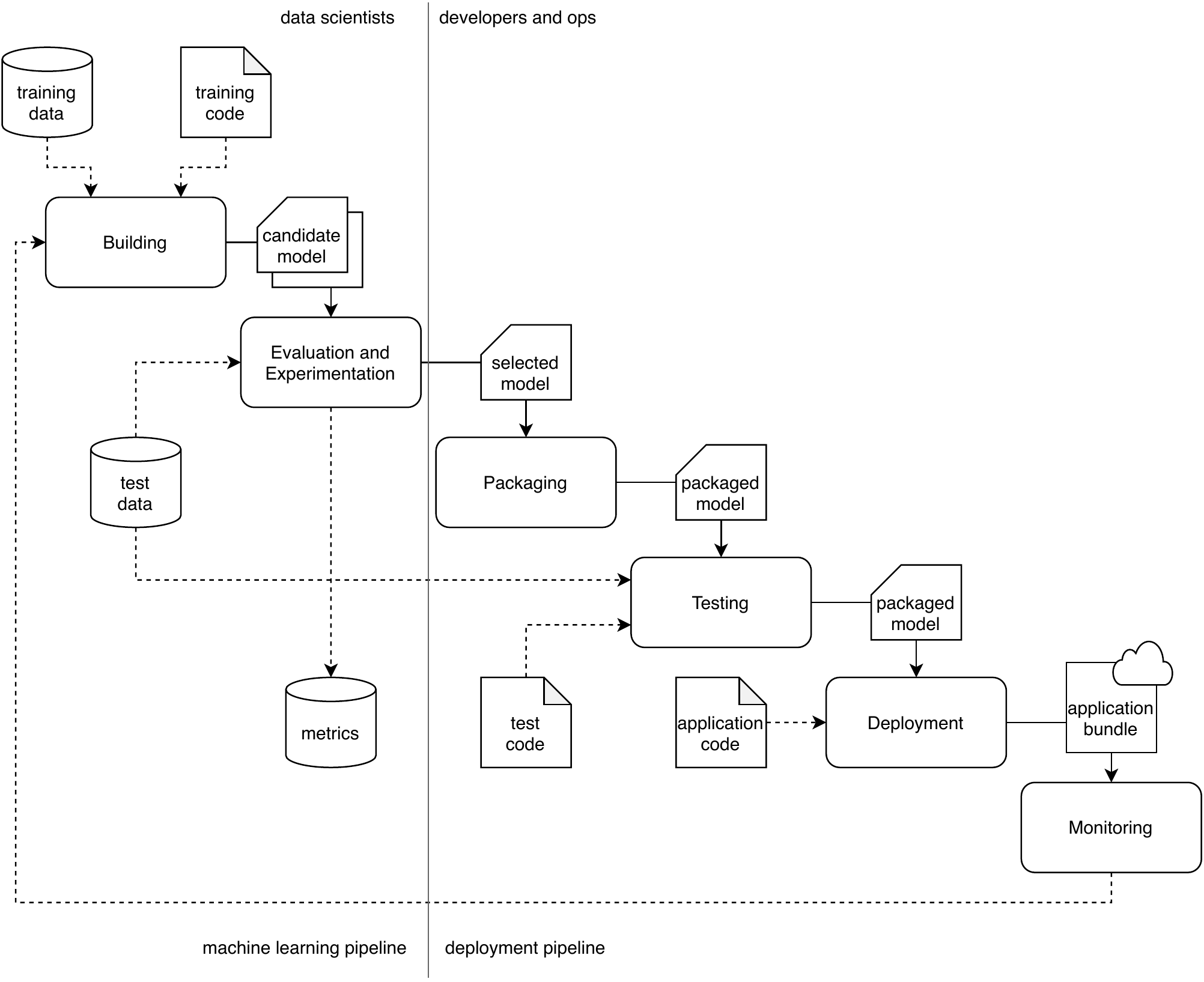}
    \caption{CD4ML pipelines and artifacts \cite{cd4ml}.}
    \label{fig:cd4ml}
\end{figure*}

\section{The Integration Challenge -- Oravizio}

%\subsection{Case Oravizio}

Oravizio\footnote{https://oraviz.io/ accessed Jan 3, 2021.} is a software product that provides data-driven information about patient-level risks related to hip (total hip arthroplasty, THA) and knee joint (total knee arthroplasty TKA) replacement surgery. In its current form, Oravizio provides three different dedicated prediction models: 
\begin{itemize}
    \item Risk of infection withing one year from surgery;
    \item Risk of revision within two years form surgery;
    \item Risk of death within two years from surgery.
\end{itemize}
With these models, Oravizio helps the collaboration and negotiation between the surgeon and a patient, so that the decisions that are taken are informed and that there is consent regarding the operation.

At the core of these models is a large amount of patient history data, collected over the years from the patients that have undergone surgery. As the number of cases is so large that no surgeon can process it manually during the appointment, this data has been used to create a risk calculation model that predicts the outcome of the surgery. The risk calculation algorithms combine manually submitted patient-specific data with comprehensive patient history data. 

Oravizio has been developed in co-operation between the two actors, a hospital that is specialized in joint replacement surgery, and Solita, a European software company headquartered in Finland. The clinical partner hospital had accumulated a large volume of data from surgeries for more than ten years, and Oravizio was based on the vision that this data asset might include factors that indicate risks for joint replacement surgeries down to an individual patient's level. 

During the development of Oravizio, the data included over 30000 patient records. Originally, the data was not organized well for post-processing. Various formats and computer systems were used, some of which had already been retired. Hence, consolidation and pre-processing were considerable tasks. The first task was to consolidate the relevant data into a data lake. Second, the data needed to be pre-processed to determine its quality and ensure its uniformity and future utility. Both of these were considerable tasks.

%\textbf{Selection of variables}.
%The prepared data included over 750 pre-operation variables from 2008 onwards. The variables included, for example,  general information (age, gender, BMI), medication, laboratory values, diagnoses, patient-reported information, and derivate variables. The clinical partner's clinical know-how was used during the research, combined with research literature and computational methods, to decide the variables to be further analyzed by data scientists. The well-documented linking between particular variables and risks in joint replacement surgery was used as a baseline for the selection. 
%The goal was to select a relatively limited set of relevant variables with a significant impact on the calculation. The modest set of input variables ensures that the finished product is practical to use, and the impact of individual risk factors can be illustrated in an easily understandable way. In the analysis, computational methods LASSO, Ridge regression, and Elastic net were used to select the most suitable parameters for the models. 

%\paragraph{Building and testing the machine learning algorithms}
Once data was uniform, several mathematical methods were considered during the research. The aim was to create an explanatory machine learning model for each risk to enable validation and ensure regulatory compliance. Luckily, the relations between the risk in joint replacement surgery and selected explanatory values were mostly known from clinical literature, so these needed no research in the process. Eventually, after estimating performance with with AUC values and ROC-curves \cite{huang2005using},  XGBoost \cite{xgboost} was selected, and the final model for the product was built accordingly. The model is deterministic by nature. Therefore, it can be validated with test data in a test environment without the need to do the validation in the final production environment.  

All development artifacts, including the scripts that were used to create ML models, were stored in the version control system (VCS), except for the ML model. The reason for this is simple: ML development was done within a scientific research project. Furthermore, the research project was performed on the premises of the clinical partner hospital. As the use of patient records is strictly regulated, the hospital's computational environment is tightly restricted and isolated. Therefore, the system was designed to be deployed in a production environment with its ML model in a \textit{locked} state, and the model handover from data scientists to software developers was done through a network drive. 

Following the same separation of concerns -- clinical model building is done in the hospital's computational environment, and deployment and operations in Solita's (Figure \ref{fig:org-boundaries}) -- the production use of Oravizio does not generate data that could be used for re-training. Instead, the needed data is generated in other clinical processes of the hospital. Due to the ML development environment's restrictive nature, frequent changes, or re-trainings to the ML models were not anticipated. However, later iterative re-training with 45000 patient records improved models' performance to a certain extent, showing the benefits of model re-training.

\begin{figure}[t]
\begin{center}
 \includegraphics[width=0.99\columnwidth]{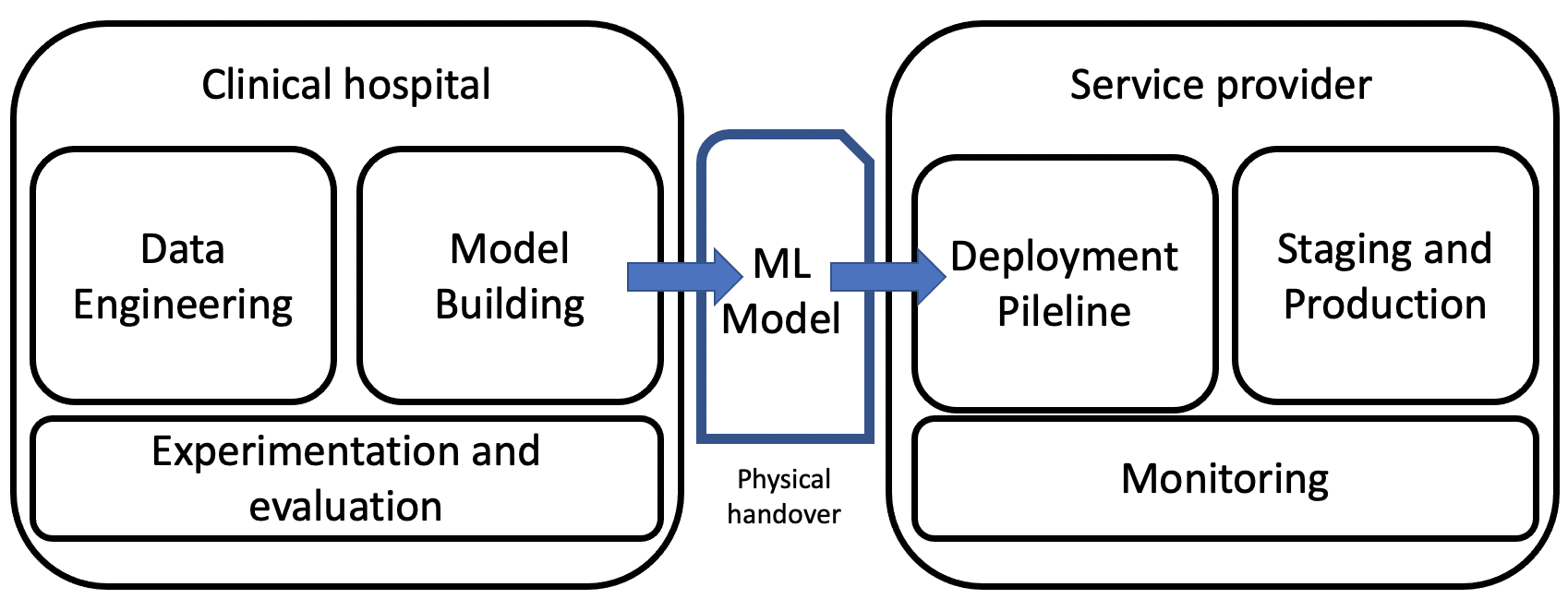}
 \caption{Organizational boundaries between clinical hospital and service provider Solita, and the physical handover of the model between organizations.}
\label{fig:org-boundaries}       % Give a unique label
\end{center}
\end{figure}

Considering the above example to demonstrate the integration challenges of an ML pipeline, we next address data sets, the model, and monitoring and related versioning.

\subsection{Data sets}

Working with data sets is the part of MLOps that is most closely associated with ownership issues. Indeed, models can originate from model libraries, and off-the-shelf systems exist for different flavors of monitoring, but data is often unique and cannot be transferred in the process. There are various reasons for this. Firstly, data sets can be so large that they cannot be easily located elsewhere. Secondly, even if it would be technically possible, the owners of the data may want to keep it to themselves. Finally, there can be other reasons, such as regulatory issues, such as in healthcare, to keep the data inside one organization.

Consequently, operations related to data seem to be the most difficult to put into practice. In general, systems like datalakes can be used to integrate data from various sources, but if amounts of data are massive and, in addition, its owners want to protect it, this option is feasible only inside one organization. 

In the case study, the situation is such that the service provider does not generate new data, but only uses the model and hence the data set that used for generating it. New data results from treating patients only, which is under the responsibility of the hospital, and not visible to the service provider. There are some downsides to this, as the approach restricts tracking down how successful the decisions were, if no operation was performed. However, the decision simplifies a lot how to manage and version data, as this is completely controlled by the hospital.

\subsection{Model} 

Based on the example, the model is a key artifact in any MLOps system. In this particular case, the model acts as the interface between hospital and service provider, who play very different roles. In fact, the model is more or less the only thing that the two organizations need to share.

Firstly, the model is created, and its quality assurance activities are carried out on the hospital's premises as a shared activity between the two organizations. The mode of operation for this is based on experiments where interesting properties are identified in the dataset, which in general is often the nature of data science projects early on \cite{aho2020demystifying}. The rhythm for the operations is defined by these experiments. If desired, the model can be re-created with more precision in given intervals or by some other valid form of meaningful iteration. Each new iteration cycle creates a new version of the model, and it may or may not be handed over to the service provider.

Secondly, the service provider is responsible for the development and the operations of the software that are necessary to use the model as basis for collaboration between the doctors and (potential) patients. The actual mode of operation is close to DevOps and continuous software engineering practices, but they have been adapted to fit regulatory requirements.
%\cite{lehtiartsu}. 
Moreover, the developers' mindset is software-focused, meaning that this mode of operation seems natural to them. In particular, they are satisfied with the given ML model, and are not interested in modifying it.

Finally, it is important to notice that the model was selected such that for a given input, it produces a deterministic output. In essence, this makes the model a deterministic software component for testing. 

\subsection{Monitoring}

As organizations responsible for Oravizio have different modes of operation, the design decisions aim at isolating them. This isolation also has an effect on monitoring the output of the model to detect bias and other problems. 

The most obvious design decision to this end is the selection of a model that produces deterministic results. Furthermore, the model does not evolve over time, but it continues producing the same results. Therefore, no drift in the results can emerge over time. Similarly, there is no need to consider retraining, except if the hospital has a new, improved data set that they wish to use to train a new model. This, however, is not detected by monitoring in this system, but requires a human decision in this case.

Finally, the tool is meant to help the doctor and the patient to discuss the risks related to a surgical operation, not to decide whether or not to perform the operation. Instead, the decision is always made by the humans, and the AI only has a supporting role in the process. Hence, the responsibility is carried by humans, not by the AI. Furthermore, in the unlikely event of the system malfunctioning and providing answers that clearly are infeasible, the doctor -- an expert in such operations -- is able to notice them and fix the situation.

\section{The Scaling Challenge -- AuroraAI}

AuroraAI initiative \cite{auroraai} is a Finnish, government-initiated program, with an objective to create the world's best public administration. Concrete means for this include linking public sector organisations to the AuroraAI network, which enables AI to facilitate their interaction with the services provided by other sectors. By removing organizational silos that complicate serving citizens in many ways, the AuroraAI network will help determine which individuals or businesses are in need of a particular service. This in turn will improve the match between users and public services while tackling inefficiency and resource waste. As the underlying framework of wellbeing, AuroraAI uses a model  proposed by Stiglitz et al. \cite{stiglitz2009report} as the baseline. The model highlights the importance of going beyond GDP or production and tackling the more difficult task of measuring people’s well-being.  Figure \ref{fig:stiglitz} addresses the multidimensional definition of people’s well-being that is based on initiatives around the world and academic research.

\begin{figure}[t]
\begin{center}
 \includegraphics[width=0.99\columnwidth]{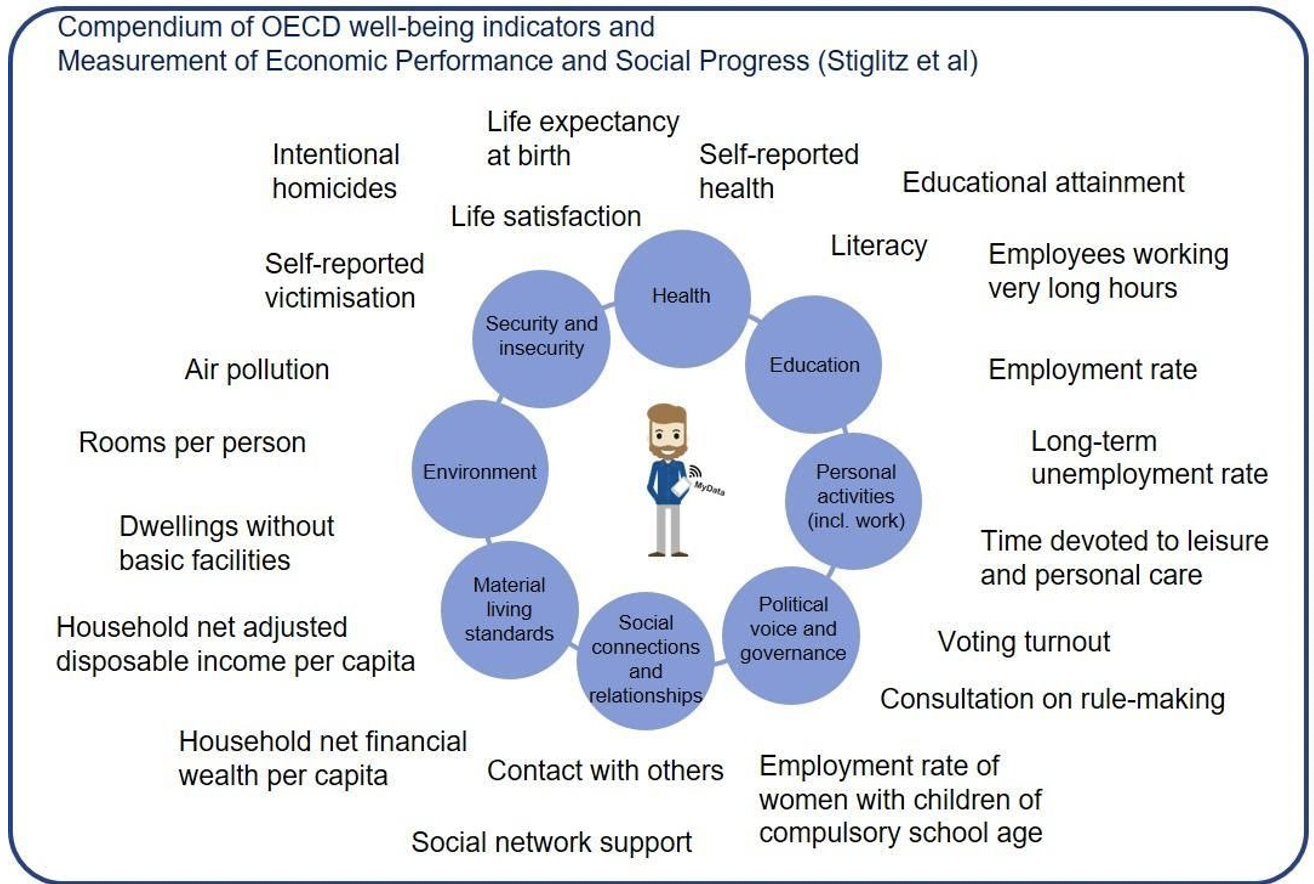}
 \caption{Multi-dimensional view to wellbeing, as adopted by AuroraAI \cite{stiglitz2009report}.}
\label{fig:stiglitz}       % Give a unique label
\end{center}
\end{figure}

Achieving the above goal requires technical conditions that enable information exchange and interoperability between different services and platforms. Seamless interaction between them will require, among other things, joint development of interfaces and communication between the development teams. Furthermore, as the goal is to rely on using AI as a part of the process, this will also mean that data sets, ML models, and monitoring -- at least in the form of governance, if nothing else -- must be considered over organizational boundaries.

%As practical means, AuroraAI has organized itself around life events to scale down the necessary data options. Hence, only a restricted set needs to be considered at this phase of the program

\subsection{Data sets}

There are several records that trace our behavior on one form or another. To protect us from a society, where all this data is available to single authority, this data is usually partitioned to several officials, based on the information they need. Hence, health records, tax information, societal data, criminal records, and so on cannot be used as a single data set, at least in Finland. Furthermore, adding data owned by private companies, such as banks, form another layer of organizational complexity in this context. 

However, it is possible to combine this data as an individual. For instance, the international non-profit organization MyData Global~\cite{mydata} is working to support collaborations among entities with interests in building a human-centric personal data ecosystem. Supporting interoperability at technical, informational and governance levels, such an ecosystem is aligned with the AuroraAI vision, where it is the individuals who combine data, not the society. The use of the digital twin paradigm \cite{glaessgen2012digital} has also been considered in this context \cite{oma-dt}, leading to citizen-level use of datasets and recommendations.

Unfortunately, such an approach, relying on datasets owned by multiple organizations, does not really provide a data set that would be easily available for ML or even deeper analysis. Firstly, MyData is not automatically shared but is something that only the individuals can release in accordance to their wishes. Secondly it is not obvious which data is true and which false, as individuals themselves provide some data, and, moreover, they can manipulate some data. 

\subsection{Model} 

In the integration challenge, the model was used as the shared component between two organizations, so that one organization trained it and released it to another without exposing its data assets. Since models are concrete assets in the ML context as well as from operations perspective, they are also something that can be easily shared in AuroraAI. However, these models are only partial, as they are built by different data owners, not based on personal data that only the citizens can construct of themselves -- different actors in the public sector only have a fragmented view. Furthermore, relying on data that people choose to share can also introduce a considerable bias.

Generalizing this approach implies that it would be possible to build systems so that ML systems are combined, following the pipes-and-filters architectural style \cite{meunier1995pipes}, with one ML system taking as input the output of another. Unfortunately, as pointed out in our recent study, this does not mean that the models would be immediately composable as such -- instead, it is considered preferrable to train a one, single model based on a combined data set than training two models for two different roles \cite{lallin-paperi}. 

Furthermore, even if it would be technically feasible to compose models, the models originating from different sources might not be compatible. For instance, while it might be meaningful from the state perspective to combine information about a citizen's wealth to health data, the models or computing systems hosting them would not be able to do this, as identity information is not included in them. Consequently, such composition would only result in statistical correlation, which might help but which might also be completely wrong, depending on the actual data. Another option is to implement a unique identity system, shared between the different information systems, so that it is clear which citizen is being attributed at a given time, and for what purpose.

For AuroraAI, this has meant that instead of aiming at automata that can provide recommendations for everyone, models are more targeted to individuals, who can use them to determine facts about their well-being. Moreover, based on the models and input from the user, recommendations are given to propose actions to add the observed well-being. Obviously, if an individual citizen chooses to share the results with municipalities, chances are that the individual in question will get a better, more targeted service proposals. However, sharing the results is by no means enforced, meaning that the resulting data set is heterogeneous from the society perspective.

\subsection{Monitoring}

As in AuroraAI, the citizens are to some extend responsible to manage their own data, much of the monitoring options should also be offered to them. However, in many of the implementations, this is tricky because the systems have been typically built from the viewpoint of the municipalities. Monitoring mechanism regarding to data access, for instance, escapes from an individual citizen, and, to gather such information, the offices responsible for hosting the data should be involved. Hence, there is a clear lack of governance related functions in today's systems. Furthermore, even if such facilities would exist, they would most likely be system specific, with little opportunities to gather unified data.

That said, individual offices often have such systems in place locally, as this is governed by law. Hence, they can monitor what takes place, and, at least to some extent, who accesses what. Opening such monitoring data to individuals with respect to their own use would probably result in increasing confidence in the use of private, personal data.

\section{Discussion}

To begin with, using AI/ML in multi-organization context, the usual integration problems emerge. APIs, data formats, performance issues and so on are similar to any other information system integration between two or more organizations. In addition, contractual and responsibility issues need to be agreed upon. To this end, using the ML model as a software component as in Oravizio, created by one party and used by another, is a way to embed ML into the realm of software development. For the purposes of the application, minimizing the integration challenge seems like a rational solution.

A new challenge in software engineering for ML is data related operations. These operations are related to the above to some extent, especially when data sets cannot be moved across data boundaries, but multiple organizations need to access the data. Moreover, data meshes and other forms of integrating data in pieces can complicate designing the more traditional parts of information systems, needed for such integration. In addition, when considering AuroraAI, it seems natural that different solutions might rely on different versions of data sets, for several reasons. For instance, it is possible that extensive data cleaning operations are needed for some applications, meaning that executing such operations frequently is impossible. Similarly, it might be so that the data must be from the same temporal range, and otherwise the operations make no sense. Similar complications are reflected to training ML models based on such data sets, as well as to monitoring how well the models work once they have been deployed. 

For operationalizing all the above in practice, the same skill gap as for starting to use MLOps within a single organization is valid -- indeed the same actions need to be taken. However, this time some of the issues are more difficult to reconcile, because the organizations may have different modes of operation and different organization cultures, as demonstrated in the Oravizio case. Moreover, restrictions, such as those related to privacy or certification, may exist on either side of the boundary, which adds an additional layer of complexity to the design. This has also been identified as a direction for future work, especially from the perspective of governance, auditing, and regulations \cite{vlad-seh2021}.

In general, to successfully perform multi-organization MLOps, we need patters of integration that help us in the process. Inspiration for these can be found from system integration \cite{hohpe2004enterprise} as well as legality patterns, proposed for open source \cite{hammouda2010open}. In fact, both solutions we have used in the examples of the paper are analogous to patterns of \cite{hammouda2010open} -- Oravizio uses the ML model as an \textit{Evaluator}, and in AuroraAI, \textit{User delegation} helps to combine data that can only be accessed by the user as a whole. The definition of such patterns remains future work, with some ideas already proposed in \cite{lallin-paperi}. 

Finally, based on both case studies reported in this paper,  it seems that if there is the will, there often is a way to create a working solution to circumvent organizational problems. The patterns described above can form a blueprint for operations, but in the end the organizations also need true collaboration in the form of joint, and possibly merged, organization-level operations (OrgOps), following the spirit of DevOps but explicitly emphasizing the importance of the culture of collaboration across organizations. For Oravizio, the organizations work in close cooperation and have shared interest in model development and maintenance in the Oravizio service; for AuroraAI, the organizations are looking over the organizational borderlines to identify best practices that could be used in other contexts, and, furthermore, some of them have executed joint initiatives for certain specific goals, such as well-being at school at a certain region.

\section{Conclusions}

There are numerous practical limitations that may prevent running MLOps pipeline within a single organization. To accommodate this situation, the MLOps pipeline needs to be split into parts that are designed and run by the respective organizations. Re-assembling the pipeline comes with the same challenges as in any IT integration project, including joint APIs, agreed data formats, and contractual obligations. In addition, due to the key characteristics of ML, data sets, ML models, and monitoring need special attention.

In this paper, we have presented two cases where ML is used in a setup where several organizations are involved. The highlighted challenges are related to integration and scaling, with real-life cases showing how ML has been addressed. As pointed out above, in both cases the teams have found practical solutions, but they are applicable to these cases only.

As a more general solution, we propose using patterns that help dealing with inter-organisation boundaries, as well as propose using joint operational mode across the organizations that are involved. However, to a large degree, this remains as a piece of future work, despite the fact we have identified such in the presented cases.

\section*{Acknowledgement}

The authors would like to thank the AuroraAI programme, Business Finland, and the members of AHMED (Agile and Holistic MEdical software Development) \& AIGA (AI Governance and Auditing) consortiums for their contribution in preparing this paper.

\bibliographystyle{plain}
\bibliography{bib}

\end{document}